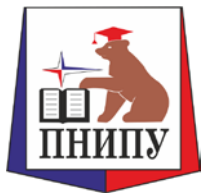

# ВЕСТНИК ПНИПУ. МЕХАНИКА
# № 2, 2020
# PNRPU MECHANICS BULLETIN
https://ered.pstu.ru/index.php/mechanics/index

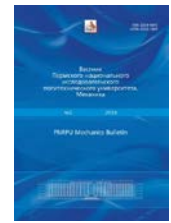

# МОДЕЛИРОВАНИЕ ПРОЦЕССОВ ФОРМИРОВАНИЯ АТОМАРНОЙ СТРУКТУРЫ СВЕРХПРОВОДЯЩЕГО СПИНОВОГО ВЕНТИЛЯ

А.В. Вахрушев[1, 2], А.Ю. Федотов[1, 2], Ю.Б. Савва[3], А.С. Сидоренко[3, 4]

[1]Удмуртский федеральный исследовательский центр Уральского отделения Российской академии наук, Ижевск, Россия
[2]Ижевский государственный технический университет имени М.Т. Калашникова, Ижевск, Россия
[3]Орловский государственный университет имени И.С. Тургенева, Орел, Россия
[4]Институт электронной инженерии и нанотехнологий имени Д. Гицу, Кишинев, Молдова

**АННОТАЦИЯ**

В работе рассматривается моделирование процессов формирования многослойного нанокомпозита, при комбинации элементов которого возникает эффект спинового вентиля. Описана актуальность и важность эффектов в области спинтроники и связанных с ними материалов и устройств. Объектом исследований являются состав и атомарная структура отдельных слоев многослойного нанокомпозита, а также состав и морфология интерфейса слоев нанокомпозита. В качестве образца анализировался образец с периодической структурой сверхпроводник – ферромагнетик, состоящей более чем из 20 чередующихся слоев ниобия и кобальта. Процесс осаждения происходил в условиях глубокого вакуума. Моделирование осуществлялось методом молекулярной динамики с использованием потенциала модифицированного метода погруженного атома. Формирование слоев производилось в стационарном режиме. Температура корректировалась при помощи термостата Нозе – Гувера. Осаждение каждой нанопленки завершалось этапом релаксации для необходимой стабилизации и перестройки структуры формируемого нанокомпозита. Рассматривалось три температурных режима осаждения: 300, 500 и 800 К. Для данных режимов выполнен анализ атомарной структуры нанопленок и переходных областей (интерфейса), образующихся между слоями. Исследование атомарной структуры нанопленок показало, что ниобий формируется кристаллическими областями различной ориентации. Для нанопленок кобальта характерно строение, близкое к аморфному. Структурные особенности интерфейса слоев сверхпроводник – ферромагнетик в значительной степени зависят от рельефа поверхности, на которую осуществляется осаждение. Наименьшую вариацию по атомарному составу имеет первая зона контакта ниобий – кобальт, так как формирование первой нанопленки происходит на ровной плоскости подложки. Анализ влияния температурного режима при формировании наносистемы свидетельствует о зависимости процессов формирования многослойных нанопленок, интерфейса нанослоев, а также состава и морфологии гетероструктур от температуры, при которой происходит изготовление нанокомпозита. Повышенная температура приводит к формированию более разреженной структуры нанослоев и увеличению зон интерфейса нанослоев за счет диффузии атомов напыляемых материалов.

© ПНИПУ

© **Вахрушев Александр Васильевич** – д.ф.-м.н, проф., e-mail: vakhrushev-a@yandex.ru, iD: 0000-0001-7901-8745.
**Федотов Алексей Юрьевич** – к.ф.-м.н., с.н.с., e-mail: alezfed@gmail.com, iD: 0000-0002-0463-3089.
**Савва Юрий Болеславович** – к.т.н., доц., e-mail: su_fio@mail.ru, iD: 0000-0002-9951-9675.
**Сидоренко Анатолий Сергеевич** – д.ф.-м.н,, проф., акад. АНМ, дир., e-mail: anatoli.sidorenko@kit.edu, iD: 0000-0001-5550-6103.

**Alexander V. Vakhrushev** – Doctor of Physical and Mathematical Sciences, Professor, e-mail: vakhrushev-a@yandex.ru,
iD: 0000-0001-7901-8745.
**Aleksey Yu. Fedotov** – CSc in Physics and Mathematics Sciences, Senior Researcher, e-mail: alezfed@gmail.com,
iD: 0000-0002-0463-3089.
**Yuri B. Savva** – CSc in Technical Sciences, Associate Professor, e-mail: su_fio@mail.ru, iD: 0000-0002-9951-9675.
**Anatoly S. Sidorenko** – Doctor of Physical and Mathematical Sciences, Professor, Academician of the ASM, Director,
e-mail: anatoli.sidorenko@kit.edu, iD: 0000-0001-5550-6103.

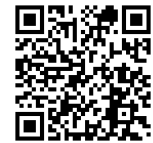

# MODELING THE PROCESSES OF ATOM STRUCTURE FORMATION OF A SUPERCONDUCTING SPIN VALVE

## A.V. Vakhrushev[1, 2], A.Yu. Fedotov[1, 2], Yu.B. Savva[3], A.S. Sidorenko[3, 4]

[1]Udmurt Federal Research Center of the Ural Branch of the Russian Academy of Sciences, Izhevsk, Russian Federation
[2]Kalashnikov Izhevsk State Technical University, Izhevsk, Russian Federation
[3]Orel State University named after I.S. Turgenev, Orel, Russian Federation
[4]Ghitu Institute of Electronic Engineering and Nanotechnologies, Chisinau, Moldova

ABSTRACT

The paper considers the modeling of a multilayer nanocomposite, the combination of elements of which gives rise to a spin valve effect. The relevance and importance of effects in the field of spintronics and related materials and devices are described. We study the composition and atomic structure of individual layers of a multilayer nanocomposite, as well as the composition and morphology of the interface of nanocomposite layers. We analyzed a sample with a periodic superconductor-ferromagnet structure consisting of more than 20 alternating layers of niobium and cobalt. The deposition process took place in a deep vacuum. The simulation was carried out by the molecular dynamics method using the potential of the modified immersed atom method. The formation of layers was carried out in a stationary mode. The temperature was adjusted using the Nose-Hoover thermostat. The deposition of each nanofilm ended with a relaxation stage for the necessary stabilization and restructuring of the formed nanocomposite. Three deposition temperature regimes were considered: 300 K, 500 K, and 800 K. For these modes, we analysed the atomic structure of nanofilms and transition regions (interface) formed between the layers. A study of the atomic structure of nanofilms showed that niobium is formed by crystalline regions of different orientations. A cobalt nanofilm is characterized by a structure close to amorphous. The structural features of the interface between the superconductor-ferromagnet layers largely depend on a relief of the surface onto which the deposition is made. The smallest variation in atomic composition is observed in the first niobium-cobalt contact zone, since the formation of the first nanofilm occurs on an even plane of the substrate. An analysis of the influence of the temperature regime during the formation of the nanosystem shows the dependence of the processes of formation of multilayer nanofilm formation, the interface of nanolayers, as well as the composition and morphology of heterostructures on the temperature at which a nanocomposite is manufactured. An increased temperature leads to the formation of a more rarefied structure of nanolayers and an increase in the zones of the interface of nanolayers due to the diffusion of atoms of the sprayed materials.

**Введение**

Развитие современных устройств микроэлектроники в настоящее время ограничено производством микросхем размером около 10 нм. Данный размер близок к физическому пределу электронных устройств из-за возникающих в них в процессе эксплуатации вихревых токов, тепловых потерь и туннельных эффектов. Существенную проблему представляет собой отвод тепла, выделяемого микросхемами при функционировании, вследствие чего ограничивается рост частоты вычислительных процессоров, а значит, и их производительности. Решить возникающие сложности может переход на новые перспективные материалы и технологии.

В последние годы стало активно развиваться такое направление наукоемких исследований, как спинтроника [1, 2]. В устройствах спинтроники вместо переноса электронов передается особая волна, называемая спиновым возбуждением, которая позволяет существенно повысить рабочую частоту процессоров [3] и увеличить плотность записи информации [4]. Квантовые эффекты благодаря взаимодействию спиновой и электронной подсистем привели к созданию таких классов перспективных материалов, как мономультиферроики [5], пьезоэлектрики [6, 7], магниторегистраторы [8, 9], термоэлектрики [10], хемоэлектрики [11] и метаструктуры [12, 13].

Одним из видов магнитных наноструктур с широким потенциалом использования являются спиновые вентили [14, 15]. Вентили состоят из нескольких магнитных пленок, разделенных магниторезистивной прослойкой. Благодаря обменному взаимодействию со смежной антиферромагнитной нанопленкой один из слоев имеет постоянную намагниченность. Для второй нанопленки направление намагниченности способно регулироваться под действием внешнего магнитного поля. Слабая связь ферромагнитных слоев влечет перестройку конфигурации магнитных моментов данных материалов в магнитных полях малой мощности, благодаря чему обеспечивается высокая чувствительность данных структур. Также возникает возможность изготовления считывающих магниторезистивных головок для жестких дисков с увеличенной плотностью записи [16]. Магнитные и электрические свойства данных многослойных материалов оптимизируют через вариацию состава, последовательности и толщины используемых слоев. Например, известно, что использование феррита кобальта в качестве магнитно-изолирующего слоя снижает эффект шунтирования и существенно увеличивает величину магнитосопротивления [17].

Несмотря на разнообразие материалов спинтроники, имеется ряд общих для них проблем, которые требуют дополнительных теоретических и экспериментальных исследований. К существенному недостатку спиновых устройств относится то, что зачастую свои эффекты они проявляют при температурах существенно ниже комнатных [18]. Постоянное охлаждение требует дополнительных затрат энергии и снижает эффективность подобных материалов, поэтому непрерывно ведется поиск элементов со стабильными спиновыми свойствами [19, 20].

Следующая важная проблема, связанная с материалами спинтроники, затрагивает реализацию технологических режимов их производства. Чаще всего многослойные наноструктуры изготавливаются при помощи лазерной абляции [21], вакуумного напыления [22] или молекулярно-лучевой эпитаксии [23]. Использование данных методов требует всестороннего изучения процессов формирования нанокомпозитов, так как даже незначительная корректировка технологических параметров может привести к изменению функциональных свойств готовых устройств. Это обусловлено нанометровыми размерами изготавливаемых образцов и сложными механизмами взаимодействия разных элементарных составов в слоях. Для анализа влияния и поиска оптимального технологического режима часто необходимы длительные экспериментальные исследования, сопряженные с затратами на дорогостоящее оборудование. Сокращение временных и экономических издержек, а также корректировка производственных процессов возможны при помощи аппарата математического моделирования.

Как отмечается в [1, 24], одно из основополагающих значений для спиновых и квантовых эффектов имеет структура контактных областей на стыке между наноплёнками. Именно взаимодействие магнитной и электронной подрешеток в слоях композита приводит к появлению магнитоэлектрического эффекта, а значит, позволяет использовать магнитное и электрическое поля в качестве управляющих. На границах раздела слоев между сверхпроводниками и ферромагнетиками возникают области конкурирующих состояний, которые под воздействием внешнего магнитного поля могут быть переведены из нормальных состояний в сверхпроводящие и обратно [25]. Кроме того, граница раздела нанослоев влияет не только непосредственно на контактный слой, но и на всю смежную область композита, изменяя его физические характеристики.

Целью данной работы (представляющую собой продолжение работы [26]) являлись теоретические исследования фундаментальных процессов формирования и физических свойств многослойных сверхпроводящих гетероструктур, включающие описание методики и комплексного анализа границы раздела слоев (интерфейса) материалов сверхпроводник – ферромагнетик, а также детальное исследование атомарной пространственной структуры и морфологии интерфейса различных нанослоев. Данное исследование является составной частью работы, направленной на идентификацию и описание топологических квантовых явлений в нанокомпозитах, которые найдут применение при разработке новых перспективных спиновых устройств.

Результаты моделирования, представленные в данной работе, являются развитием выполненных ранее исследований процессов осаждения нанослоев и наноэлементов на сплошные [27, 28] и пористые [29] поверхности твердого тела.

## 1. Математическая модель и теоретические основы моделирования

Исследование процессов формирования и структуры многослойных материалов спинтроники осуществляется как экспериментальными, так и теоретическими методами. Теоретические модели, как правило, основаны на феноменологическом подходе [24, 30], что делает невозможным получение детальной информации о механизмах взаимодействия отдельных атомов, наноструктур, наноплёнок. Также возникают сложности с наблюдением за динамическими процессами модификации исследуемых объектов.

Данных недостатков лишен математический аппарат молекулярной динамики. Она описывает движение каждого атома наносистемы в определенный момент времени, поэтому способна воспроизводить эволюцию наноэлементов и их свойств. Основу метода составляют уравнения движения всех атомов наносистемы, дополненные начальными условиями в виде координат и скоростей атомов:

$$m_i \frac{d^2 \mathbf{r}}{dt^2} = -\frac{\partial U(\mathbf{r})}{\partial \mathbf{r}_i} + \mathbf{F}_i^{ex}, \quad \mathbf{r}_i(t_0) = \mathbf{r}_{i0},$$
$$\frac{d\mathbf{r}_i(t_0)}{dt} = \mathbf{V}_{i0}, \quad i = 1, \ldots N, \quad (1)$$

где $N$ – количество атомов системы; $m_i$ – масса атома; $\mathbf{r}_{i0}, \mathbf{r}_i(t)$ – начальный и текущий радиус-векторы; $U(\mathbf{r})$ – потенциал системы, который зависит от взаимного расположения всех атомов; $\mathbf{V}_{i0}, \mathbf{V}_i(t)$ – векторы скорости в начальный и текущий момент соответственно; $\mathbf{r}(t) = \{\mathbf{r}_1(t), \mathbf{r}_2(t), .., \mathbf{r}_K(t)\}$ – переменная, указывающая на зависимость от координат всех атомов; $\mathbf{F}_i^{ex}$ – внешняя сила, отражает взаимодействие наносистемы с внешней средой, в том числе отвечает за корректировку энергии для поддержки постоянной температуры.

Метод молекулярной динамики опирается на понятие потенциала $U(\mathbf{r})$, который отвечает за характер и величину взаимодействий атомов наносистемы. Вид потенциала может быть различен, но в последнее время благодаря своей точности и адекватности большую популярность получили многочастичные силовые поля. В данной работе использовался потенциал модифицированного метода погруженного атома – MEAM (Modified Embedded Atom Method). Виды и особенности большого числа современных многочастичных потенциалов хорошо описаны в обзорах [31, 32].

Модифицированный метод погруженного атома выведен при помощи теории функционала плотности – DFT (Density Functional Theory). В данном методе результирующий потенциал наносистемы представляется в виде суммы вкладов энергий отдельных атомов, причем отдельно выделяется вклад парного и многоэлементного взаимодействия

$$U(r) = \sum_i U_i(r) = \sum_i \left( F_i(\overline{\rho}_i) + \frac{1}{2} \sum_{j \neq i} \varphi_{ij}(r_{ij}) \right), \quad (2)$$
$$i = 1, 2, ..., N,$$

где $U_i(r)$ – потенциал отдельного атома, влияющий на характер и степень взаимодействия в уравнениях движения (1); $F_i$ – функция погружения атома, зависит от электронной фоновой плотности $\overline{\rho}_i$; $\varphi_{ij}(r_{ij})$ – вклад парного потенциала в общую энергию, изменяется с удалением на расстояние $r_{ij}$.

Функция погружения корректирует силовое поле, созданное парными взаимодействиями, и уточняет его значение. Данная величина обусловлена наличием электронного газа в материале и в соответствии с [33, 34] может быть вычислена по формуле

$$F_i(\overline{\rho}_i) = \begin{cases} A_i E_i^0 (\overline{\rho}_i) \ln(\overline{\rho}_i), & \overline{\rho}_i \geq 0, \\ -A_i E_i^0 \overline{\rho}_i, & \overline{\rho}_i < 0, \end{cases} \quad (3)$$

где $A_i$ – эмпирический параметр силового поля; $E_i^0$ – значение энергии сублимации; $\overline{\rho}_i$ – величина фоновой электронной плотности.

Для вычисления фоновой электронной плотности в точке погружения используется следующее выражение, где свои вклады вносят все электронные орбитали атомов различной конфигурации:

$$\overline{\rho}_i = \frac{\rho_i^{(0)}}{\rho_i^0} G(\Gamma_i), \quad \Gamma_i = \sum_{k=1}^{3} t_i^{(k)} \left( \frac{\rho_i^{(k)}}{\rho_i^{(0)}} \right)^2, \quad (4)$$

где параметры $t_i^{(k)}$ являются весовыми коэффициентами модели; $\rho_i^0$ – величина фоновой электронной плотности исходной структуры; $\rho_i^{(k)}$ – характеризуют изменение электронной плотности в реальных условиях. Индексы $k = 1, 2, 3$ принадлежат различным видам электронных орбиталей рассматриваемого атома.

Выделяют сферически симметричную одноэлектронную $s$-орбиталь и угловые электронные $p$-, $d$-, $f$-облака. Для каждой орбитали используется свое выражение, в соответствии с которым вычисляется ее электронная плотность распределения:

$s$-орбиталь:

$$\rho_i^{(0)} = \sum_{i \neq j} \rho_j^{A(0)}(r_{ij}) S_{ij}, \quad (5)$$

$p$-орбиталь:

$$\left(\rho_i^{(1)}\right)^2 = \sum_\alpha \left[ \sum_{i \neq j} \frac{r_{ij\alpha}}{r_{ij}} \rho_j^{A(1)}(r_{ij}) S_{ij} \right]^2, \quad (6)$$

$d$-орбиталь:

$$\left(\rho_i^{(2)}\right)^2 = \sum_{\alpha,\beta} \left[ \sum_{i \neq j} \frac{r_{ij\alpha} r_{ij\beta}}{r_{ij}^2} \rho_j^{A(2)}(r_{ij}) S_{ij} \right]^2 - \frac{1}{3} \left[ \sum_{i \neq j} \rho_j^{A(2)}(r_{ij}) S_{ij} \right]^2, \quad (7)$$

$f$-орбиталь:

$$\left(\rho_i^{(3)}\right)^2 = \sum_{\alpha,\beta,\gamma} \left[ \sum_{i \neq j} \frac{r_{ij\alpha} r_{ij\beta} r_{ij\gamma}}{r_{ij}^3} \rho_j^{A(3)}(r_{ij}) S_{ij} \right]^2 - \frac{3}{5} \sum_\alpha \left[ \sum_{i \neq j} \frac{r_{ij\alpha}}{r_{ij}} \rho_j^{A(3)}(r_{ij}) S_{ij} \right]^2, \quad (8)$$

где $\rho^{A(h)}$ – радиальные функции; $S_{ij}$ – экранирующая функция потенциала; $r_{ij\alpha}$ – компонента $\alpha$ из вектора расстояния между атомами, $\alpha, \beta, \gamma = x, y, z$.

Функционал $G(\Gamma)$ в выражении (4) может быть задан различными способами. Одна из наиболее популярных и удобных формулировок записывается в виде зависимости

$$G(\Gamma) = \begin{cases} \sqrt{1+\Gamma}, & \Gamma \geq -1, \\ -\sqrt{|1+\Gamma|}, & \Gamma < -1. \end{cases} \quad (9)$$

Весовые коэффициенты модифицированного метода погруженного атома из (4) также имеют аддитивную связь с одноэлектронными радиальными функциями

$$t_i^{(k)} = \frac{\sum_{i \neq j} t_{0,j}^{(k)} \rho_j^{A(0)} S_{ij}}{\sum_{i \neq j} \left(t_{0,j}^{(k)}\right)^2 \rho_j^{A(0)} S_{ij}}, \quad (10)$$

где $t_{0,j}^{(k)}$ – эмпирические параметры, которые зависят от химического типа элемента рассматриваемого атома.

Сглаживания энергии взаимодействия на расстоянии в MEAM добиваются введением функции экранирования. При помощи функции экранирования затухание потенциала происходит постепенно, что позволяет физически более точно описать свойства наноматериалов и снизить вычислительные затраты при проведении моделирования:

$$S_{ij} = f_c\left(\frac{r_c - r_{ij}}{\Delta r}\right) \prod_{k \neq i, j} f_c\left(\frac{C_{ikj} - C_{\min,ikj}}{C_{\max,ikj} - C_{\min,ikj}}\right), \quad (11)$$
$$C_{ikj} = 1 + 2\frac{r_{ij}^2 r_{ik}^2 + r_{ij}^2 r_{jk}^2 - r_{ij}^4}{r_{ij}^4 - \left(r_{ik}^2 - r_{jk}^2\right)^2},$$

$$f_c(x) = \begin{cases} 1, & x \geq 1 \\ \left[1 - (1-x)^4\right]^2, & 0 < x < 1, \\ 0, & x \leq 0 \end{cases} \quad (12)$$

где $C_{\min}, C_{\max}$ – параметры взаимного влияния атомов, в зависимости от их химических типов выставляются для каждой тройки атомов с номерами $i, j, k$; $r_c$ – расстояние, на котором происходит обрезание силового поля; $\Delta r$ – параметр, превышающий расстояние обрезания, используется для сглаживания потенциала.

## 2. Постановка задачи и программный комплекс

Задача исследования процессов формирования и изучения структуры материалов сверхпроводник – ферромагнетик решалась для многослойного композита на основе кобальта и ниобия. Данный композит является функциональным материалом со спиновым вентилем и используется в устройствах с управляемой эффективной энергией обмена и переключателями энергии [35, 36]. Композит представляет собой многослойную структуру, состоящую из ферромагнитных нанопленок кобальта, которые разделены тонкими сверхпроводниками ниобия. Общая схема гетероструктуры (повернутая на 90°) продемонстрирована на рис. 1.

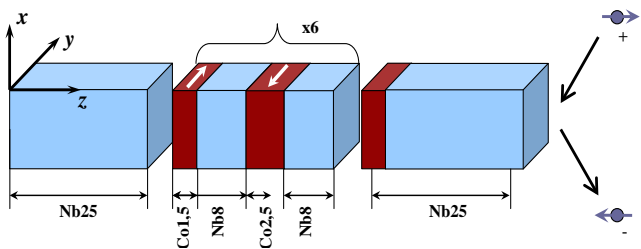

Рис. 1. Структурная схема многослойного композита кобальта и ниобия со спиновым вентилем

Fig. 1. The block diagram of the multilayer composite of cobalt and niobium with the spin valve

Важность материалов со спиновыми эффектами на основе кобальта уже отмечалась в [17]. Цифрами рядом с элементами в слоях представлена их толщина в нанометрах. Изготовление образца осуществляется через осаждение материала в вакууме. В общем случае композит содержит около 20 слоев. Тем не менее процессы формирования наноматериала, а также его структурные особенности в слоях схожи между собой. Поэтому в данной работе рассматривается осаждение только первых трех нанопленок кобальт – ниобий.

Первый слой материала, образованный атомами ниобия, является подложкой и основой для вакуумного напыления последующих нанопленок. Подложка помещается в нижнюю область расчетной ячейки, крайний ее слой фиксируется, чтобы исключить хаотичное перемещение образца в процессе моделирования. Начальные координаты атомов подложки задаются исходя из ее кристаллической структуры. Скорости атомов устанавливаются в соответствии с распределением Максвелла, соответствующим начальной температуре. На границах расчетной ячейки накладываются периодические граничные условия. На поверхности, через которую осуществляется напыление, заданы граничные условия отражения, чтобы осаждаемые атомы не покидали систему моделирования. Процесс напыления моделируется посредством введения атомов в зоне над подложкой. При этом осаждаемым атомам придается определенная скорость по направлению к подложке. Напыление слоев происходит поэтапно. Магнитное поле в наносистеме отсутствует.

Осаждение каждой нанопленки завершалось этапом релаксации для необходимой стабилизации и перестройки атомарной структуры формируемого нанокомпозита. На стадии релаксации атомы наносистемы перестраиваются и самоорганизуются в более выгодные энергетические позиции при постоянной температуре в отсутствие внешних воздействий. Управляемым параметром при этом является температура релаксации. Стабилизация положения атомов свидетельствуют об их устойчивом состоянии.

В качестве программного средства для проведения теоретических исследований использовался вычислительный комплекс LAMMPS (Large-scale Atomic/Molecular Massively Parallel Simulator) [37]. Данный программно-инструментальный пакет является свободно распространяемым, содержит возможности для выполнения параллельных вычислений и поддерживает разноуровневые математические модели, в том числе молекулярную динамику. Алгоритмы анализа результатов описывались на языке tcl и C++. На основе LAMMPS были разработаны и реализованы скрипты и алгоритмы для детального исследования структуры материалов сверхпроводник – ферромагнетик и определения их пространственного профиля.

Математическая модель и программный комплекс находятся в тесном взаимодействии с реальными функциональными наноматериалами, так как позволяют изучить возможности оптимизации их свойств и повысить эффективность технологии их изготовления. Как пока-зано на рис. 2, обмен информацией между математической моделью и реальными наноструктурами происходит через входные данные и процессы управления.

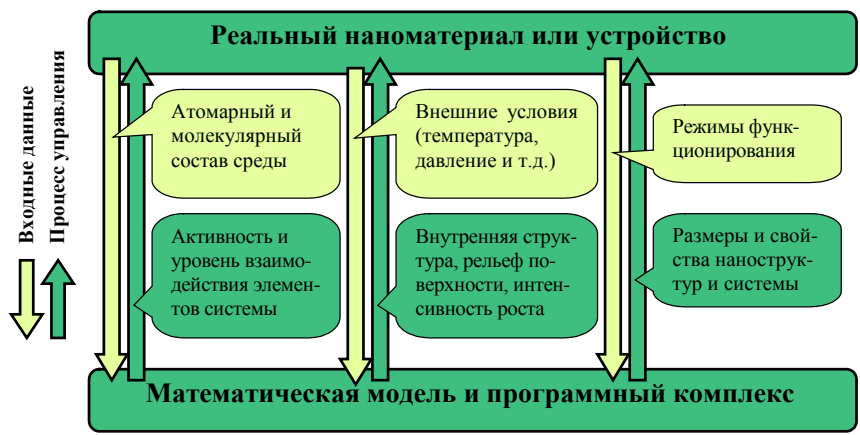

Рис. 2. Обмен информацией между моделью, программным комплексом и реальным материалом

Fig. 2. Information exchange between a model, a software package and real material

Влияние таких входных данных, как атомарный и молекулярный состав среды, термодинамические условия и режимы функционирования через модели и алгоритмы управления, учитывают и корректируют активность и уровень взаимодействия элементов, внутреннюю атомарную структуру, рельеф поверхности, интенсивность роста слоев нанокомпозита, размеры и свойства наноструктуры и наносистемы в целом.

**3. Результаты моделирования**

Формирование многослойных нанокомпозитов осуществлялось в несколько стадий. На первой стадии производилось осаждение кобальта на подложку, образованную атомами ниобия. Подложка имела кристаллическую структуру высотой 3,7 нм и шириной 13,2 нм в горизонтальных направлениях. Отметкой нулевой высоты, с которой начинали формироваться слои осаждаемого материала, являлась поверхность подложки. Отсчет высоты от поверхности подложки связан с удобством представления графиков. Число атомов ниобия в подложке соответствовало значению 33,6 тыс.

Первая серия вычислительных экспериментов была направлена на исследование процессов формирования наноструктур при комнатной температуре – 300 К. Температура в наносистеме контролировалась при помощи термостата Нозе – Гувера [38]. Термостатом поддерживалась температура подложки. Осаждаемые атомы имели направленную скорость, поэтому в непосредственной корректировке термостатом не участвовали.

Реальная гетероструктура из ниобия и кобальта со спиновым вентилем, предложенная в [35], имеет размерные характеристики, продемонстрированные на рис. 1. Для соответствия результатов моделирования и экспериментальных данных и формирования нанопленок требуемой целевой толщины на подложку на первой стадии осаждалось 18 тыс. атомов кобальта, на второй – 70 тыс. атомов ниобия и на третьем этапе напылялись снова 30 тыс. атомов кобальта. В результате были сформированы три однокомпонентные нанопленки высотой 1,5; 8,0 и 2,5 нм. Изображение многослойного образца, полученного в результате математического моделирования, проиллюстрировано на рис. 3.

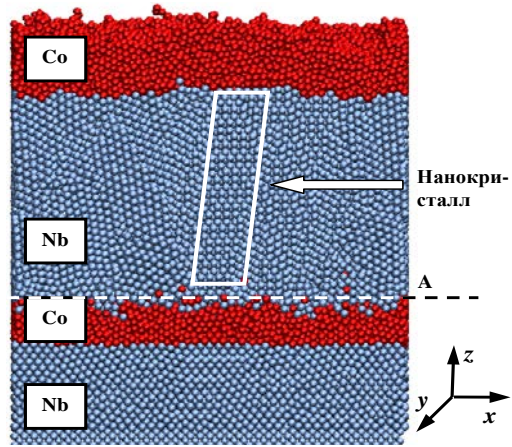

Рис. 3. Многослойная гетероструктура ниобия и кобальта, полученная моделированием при температуре 300 К

Fig. 3. Multilayer heterostructure of niobium and cobalt obtained by simulation at a temperature of 300 K

Изображения на рис. 3 хорошо характеризуют качественную картину процессов формирования гетероструктуры ниобия и кобальта и визуальную структуру слоев. Строение слоев, образованных атомами ниобия, близко к кристаллическому. При этом группы атомов объединяются в нанокристаллиты с разной пространственной ориентацией. На рис. 3 выделен один нанокристалл. Нанопленки кобальта имеют аморфную структуру. Заметна размытость контактной области между слоями и менее ровный, по сравнению с ниобием, профиль поверхности. Данные о рельефности полученных нанопленок кобальта подтверждаются пространственной картой высот, построенной в горизонтальном направлении и приведенной на рис. 4. Расположение про-филя поверхности нанокомпозита (А) показано на рис. 3 пунктирной линией. Распределение высот слоя кобальта представлено до начала осаждения следующей нанопленки на материал.

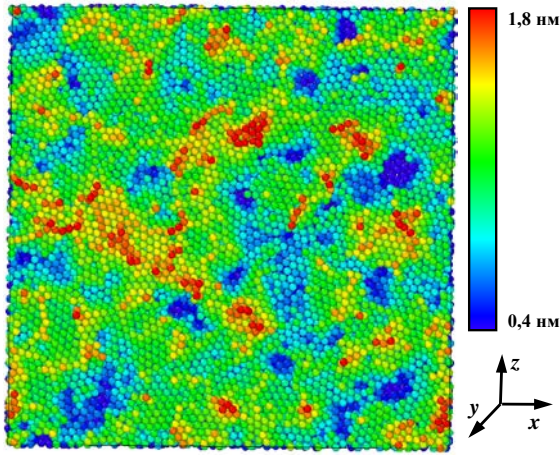

Рис. 4. Горизонтальный профиль поверхности (А) нанопленки кобальта, полученный при температуре 300 К

Fig. 4. The horizontal surface profile (A) of the cobalt nanofilm obtained at a temperature of 300 K

Анализ распределения высот, продемонстрированного на рис. 4, свидетельствует об имеющемся рельефе и неровностях сформированной поверхности. Наблюдается перепад высоты в диапазоне от 0,4 до 1,8 нм с приблизительно равномерно расположенными участками углублений и возвышений. При этом большая площадь поверхности имеет среднестатистическое значение высоты около 1,5 нм. Перепады в профиле контактного слоя в дальнейшем будут отражены в изменениях послойного состава исследуемого нанокомпозита. Для поверхности нанопленок ниобия был характерен меньший разброс высот и наличие более равномерно распределенного строения.

Визуальные характеристики и строение многослойных наноструктур сверхпроводник – ферромагнетик дополнены количественным описанием состава конечного образца, сформировавшегося к моменту времени моделирования. Распределение атомов в вертикальном направлении нанокомпозита представлено на рис. 5.

Анализ структуры осуществлялся поочередно в горизонтальных слоях толщиной 0,1 нм, поэтому на графиках приведена процентная доля атомов ниобия и кобальта по отношению к общему числу частиц на текущем участке. Из-за значительной протяженности образца и для большей информативности на промежутке высоты 3,5–7,5 нм на графиках сделан разрыв, показанный на рис. 5 наклонными линиями. Изменение состава в указанном диапазоне высот незначительно.

В результате осаждения трех нанопленок с поочередно чередующимся составом в нанокомпозите были сформированы три контактные области слоев сверхпроводника и ферромагнетика. Зонам контакта на рис. 5 соответствуют области изменения состава и строения, которые происходят на высотах 0,0 нм (верхняя плоскость подложки), 1,5 нм и 9,5 нм. Данные высоты совпадают с толщинами первых нескольких нанопленок ниобия и кобальта, обладающих спиновым вентилем и описанных в [35]. Первая область контакта ниобий – кобальт на высоте 0,0 нм имеет незначительную размытость. Формирование нанопленки в данном случае происходит на ровной поверхности подложки, поэтому взаимного проникновения атомов между слоями и перемешивания состава не наблюдается.

Осаждение второй и третьей нанопленок происходит на уже сформированные на предыдущем этапе моделирования рельефные поверхности. Их рост сопровождается более размытыми контактными областями. Наличие смешанного состава обусловлено проникновением осаждаемых атомов внутрь слоев ранее образованных нанопленок. Проникновение может быть как одномоментным при непосредственном контакте с поверхностью, так и отложенным, связанным с перестроениями атомов вследствие тепловых флуктуаций и стремления занять более энергетически выгодное положение. Как видно на рис. 5, при температуре моделирования 300 К изменение состава в нанокомпозите осуществляется плавно, без резких скачков и перепадов.

В работе была проведена серия вычислительных экспериментов с аналогичными составами, где формирование наноструктур производилось при повышенных температурах 500 К и 800 К.

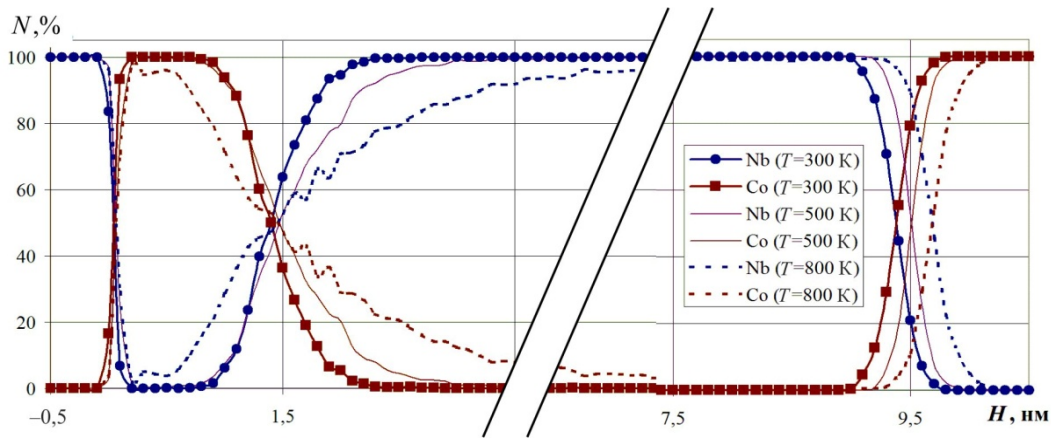

Рис. 5. Относительный послойный состав нанокомпозита Nb–Co

Fig. 5. The relative layered composition of the nanocomposite Nb-Co

Полученные результаты моделирования имеют во многом схожую структуру и визуально мало отличаются от ранее приведенного изображения образцов, представленного на рис. 3. Для количественных оценок строения материалов в режимах повышенных температур имеются некоторые различия. На рис. 5 графики состава нанокомпозита при 800 К построены пунктирной линией. Как видно, рост температуры привел к увеличению толщин формируемых нанопленок, а значит, к уменьшению их плотности. В суммарном размере высота всех слоев выросла на 0,3 нм. Кроме того, с повышением температуры увеличились области смешанного состава, расположенные на контактных границах. Поведение графиков стало носить более изменчивый характер, заметны некоторые вариации состава. Проведенные исследования свидетельствуют о существенной зависимости свойств, структуры и состава нанокомпозита от температуры, при которой осуществляется изготовление наноматериалов и устройств со спиновыми эффектами.

Визуальный анализ состава и структуры контактных областей может быть проведен на основе профиля поверхности, показанного на рис. 4, и общего строения нанокомпозита, приведенного на рис. 3. Для нанопленок ниобия характерно образование кристаллических областей с разной пространственной ориентацией. Формирование кристаллитов происходит вследствие релаксационных механизмов, возникающих внутри материала при перестроениях атомов. Слои кобальта имели строение, более близкое к аморфному состоянию, четко выраженных кристаллических участков не наблюдалось.

Для идентификации кристаллической и аморфной структуры материала в теоретических исследованиях используется параметр идеальности решетки $C_S$, алгоритм вычисления и механизм использования которого подробно описаны в [39]. Параметр идеальности определяется через отклонение между центральным и соседними, близко расположенными атомами для числа ближайших соседей материала. Полученное значение нормируется по отношению к квадрату параметра кристаллической решетки. Для идеальных кристаллов величина $C_S$ приближается к нулю, для других материалов имеет положительное значение. Изменение параметра идеальности решетки в слоях нанокомпозита ниобий – кобальт представлено на рис. 6. Построенный график изменения параметра идеальности решетки в слоях нанокомпозита подтверждает ранее сделанные выводы о строении его слоев. Нанопленки ниобия имеют меньшее значение параметра, чем наноструктуры кобальта. Величина $C_S$ на уровне 0,4 говорит о кристаллическом строении материала с присутствующими дефектами решетки. Для аморфных веществ параметр имеет положительное значение больше единицы. В связи с этим структура слоев кобальта близка к аморфному состоянию. Повышение температуры формирования нанокомпозита приводит к разупорядочиванию положений атомов, а следовательно, к увеличению параметра идеальности решетки. При температуре 500 К в нанопленке ниобия наблюдаются скачки значения $C_S$, а при 800 К происходит его существенное увеличение в данной области. Поведение структуры подложки при всех рассматриваемых температурах оставалось стабильным. Значительное снижение параметра идеальности решетки на крайнем участке последней нанопленки обусловлено принципом его вычисления вблизи свободной поверхности.

**Выводы и заключение**

В работе описаны математическая модель, постановка задачи и программно-ориентированный комплекс с адаптированными алгоритмами анализа для детализации процессов формирования многослойных нанокомпозитов. Методика теоретических исследований направлена на изучение атомарного состава и строения слоев и нанопленок, атомарной структуры и упорядоченности областей интерфейса слоев, установления механизмов роста наноструктур в зависимости от различных температур изготовления наноматериалов.

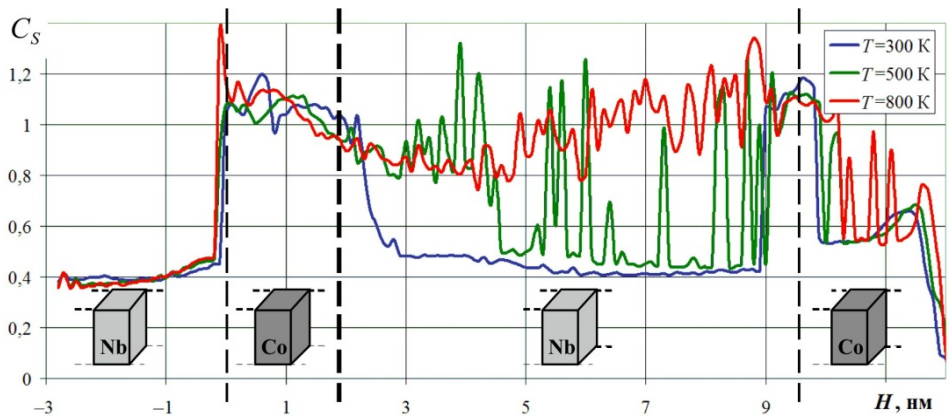

Рис. 6. Изменение параметра идеальности решетки в слоях нанокомпозита Nb–Co

Fig. 6. Change in the lattice ideality parameter in the Nb-Co nanocomposite layers

Показано, что при формировании нанокомпозитов ниобия и кобальта методом поэтапного многослойного напыления в условиях высокого вакуума атомарная структура материала зависит от состава слоев. Для нанопленок ниобия характерно образование кристаллических областей с разной пространственной ориентацией. Слои кобальта имели строение, более близкое к аморфному состоянию, четко выраженных кристаллических участков не наблюдалось. Данные о структуре нанопленок количественно подтверждены графиками изменения нормированного параметра идеальности решетки. Нанопленки ниобия имеют значение параметра на уровне 0,4, что говорит о кристаллическом строении материала с присутствующими дефектами решетки. Параметр идеальности решетки для слоев кобальта близок к 1,0, что характерно для аморфного вещества.

Особенности структуры интерфейса между нанопленками ниобия и кобальта в значительной степени определяются рельефом поверхности, на которую происходит осаждение материала. Напыление на ровную поверхность приводит к ярко выраженной границе слоев. Искаженный профиль высот обусловливает размытость контактных зон. Наименьшую протяженность имеет первая контактная область ниобий – кобальт, так как ее формирование происходит на ровной плоскости подложки. Максимальный перепад высот в диапазоне от 0,4 до 1,8 нм с равномерно расположенными участками углублений и возвышений наблюдается в профиле поверхности между первой и второй нанопленками.

Анализ термодинамических режимов формирования нанокомпозита свидетельствует о существенной зависимости свойств, структуры и состава материала от температуры в наносистеме. Увеличение температуры подложки привело к росту толщин формируемых нанопленок и уменьшению их плотности. С повышением температуры увеличиваются области смешанного атомарного состава, расположенные на границах слоев. Наблюдаются заметные вариации атомарного состава и разупорядочивание положений атомов в узлах решетки, что приводит к переходу некоторых участков нанопленок из кристаллического состояния в аморфное.

Проведенные исследования позволяют получить детальную информацию о строении и атомарном составе нанокомпозита ниобий – кобальт, механизмах формирования наноструктур, нанопленок и областей интерфейса слоев при различных температурных режимах изготовления наноматериала. Полученные по результатам моделирования данные могут быть использованы при определении параметров программы экспериментальных исследований и для корректировки и оптимизации параметров технологических процессов формирования многослойных наносистем и устройств на их основе со спиновыми эффектами.

**Благодарности**

Исследование выполнено при финансовой поддержке проекта 0427-2019-0029 УрО РАН «Исследование закономерностей формирования и расчет макропараметров наноструктур и метаматериалов на их основе методами многоуровневого математического моделирования» (раздел статьи 1); проекта РНФ 20-62-47009 «Физические и инженерные основы вычислителей не фон Неймановской архитектуры на базе сверхпроводниковой спинтроники» (разделы статьи 2, 3).

**Acknowledgments**

The study was supported by the Project No. 0427-2019-0029 Ural Branch of the Russian Academy of Sciences Study of the laws of formation and calculation of macroparameters of nanostructures and metamaterials based on them using multilevel mathematical modeling (Section of Article 1); the Russian Science Foundation, Project No. 20-62-47009 Physical and engineering fundamentals of calculators not von Neumann architecture based on superconducting spintronics (Sections of Article 2,3).